\documentclass[conference]{IEEEtran}
\usepackage{graphicx}
\usepackage{cite}
\usepackage{epsfig}
\usepackage{amsmath}
\usepackage[center]{caption}
\usepackage{tabularx,booktabs}
\usepackage{comment}

\usepackage{soul}  %% help text color & highlight color
\usepackage[table,xcdraw]{xcolor}

\begin{document}

\title{Modeling and Analysis of Analog Non-Volatile Devices for Compute-In-Memory Applications }
\author{Carl Brando, Minseong Park, Sayma Nowshin Chowdhury,\\ Matthew Chen, Kyusang Lee and Sahil~Shah
\thanks{Carl Brando, S. Chowdhury, M Chen and S. Shah are with the Department
of Electrical and Computer Engineering at University of Maryland, College Park, MD, USA e-mail: (sshah389@umd.edu). M park and K Lee are with the Department
of Electrical and Computer Engineering at University of Virginia, Charlottesville, VA, USA e-mail: (kl6ut@virginia.edu).}% <-this % stops a space}
}

\IEEEoverridecommandlockouts
\markboth{Journal of \LaTeX\ Class Files,~Vol.~14, No.~8, August~2015}%
{Shell \MakeLowercase{\textit{et al.}}: Bare Demo of IEEEtran.cls for IEEE Journals}

\maketitle

\begin{abstract} 
This paper introduces a novel simulation tool for analyzing and training neural network models tailored for compute-in-memory hardware. The tool leverages physics-based device models to enable the design of neural network models and their parameters that are more hardware-accurate. The initial study focused on modeling a CMOS-based floating-gate transistor and memristor device using measurement data from a fabricated device. Additionally, the tool incorporates hardware constraints, such as the dynamic range of data converters, and allows users to specify circuit-level constraints. A case study using the MNIST dataset and LeNet-5 architecture demonstrates the tool's capability to estimate area, power, and accuracy. The results showcase the potential of the proposed tool to optimize neural network models for compute-in-memory hardware.

%analysis and optimization for using analog non-volatile memory elements, specifically Resistive Random Access Memory (ReRAM) and Floating-Gate (FG) transistors. The work incorporates the non-linear behavior of the memory devices during training and inference. 

\end{abstract}
\IEEEpeerreviewmaketitle

\section{Introduction}
Machine learning algorithms are increasingly being used to analyze data generated by edge devices, such as autonomous robots \cite{mcguire_minimal_2019} and remote sensors\cite{olsson_event_2016}. The complexity of these algorithms has led to a growing need to research and develop unique computing architectures, circuits, and devices. Specifically, for edge applications with constrained resources, there is a significant need for developing systems that can enable high energy efficiency and small area while sustaining the required accuracy. 

This need has led to a variety of system architectures that can reduce the overall power consumption of machine learning accelerators. Specifically, studies have used software-driven optimizations to reduce the overall model size \cite{han_learning_nodate}, leveraged data flow between memory and compute in neural networks to develop optimal digital architectures \cite{sze_designing_2017}, and alternative architectures for performing computation \cite{chowdhury_hardware_2022}. Compute-in-memory architectures are a promising direction that has been shown to reduce the overall power consumption of workloads that are memory bound, which is the case for machine learning algorithms \cite{ankit_puma_2019}. Various traditional memory devices such as SRAM \cite{ali_imac_2020} and DRAM \cite{mutlu_processing_2019} as well emerging memory devices such as spintronics \cite{fan_leveraging_2017,sengupta_hybrid_2016} and phase-change memory \cite{khaddam-aljameh_hermes_2021} devices have been shown to be used for compute-in-memory architectures. 

\begin{figure}[ht]
  \begin{center}
  \includegraphics[ width=0.8\columnwidth]{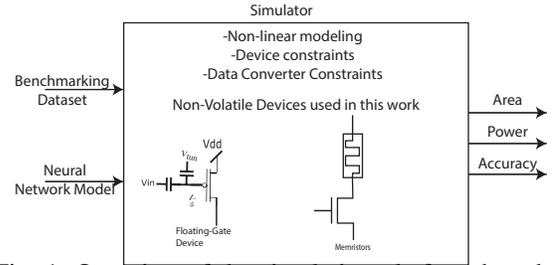}
  \vspace{-5mm}
  \end{center}
  \caption{Overview of the simulation platform based on measured non-volatile devices.}\label{fig:overview}
  \vspace{-5mm}
\end{figure}

Compute-in-memory architecture can greatly benefit from analog non-volatile devices. Analog non-volatile devices can provide a greater degree of density for storing weights with similar precision compared to traditional memory devices. However, the current implementation of compute-in-memory architectures that use analog non-volatile devices have a fixed system architecture and are benchmarked against specific datasets \cite{ali_imac_2020}. Further, a significant number of studies use a separate training framework and then in turn tune the weights for the analog synapses \cite{tiled_reram}. However, given the non-linearity and variations observed in analog non-volatile devices scaling this framework to a large number of devices is challenging. 

This study introduces our preliminary effort in developing a Python-based simulation framework designed to explore the extensive design space of analog non-volatile device-based compute-in-memory architectures and evaluate hardware performance across diverse datasets and neural network models. Additionally, the framework accurately models analog non-volatile devices by measuring fabricated devices. Currently, the simulator encompasses two such devices: Resistive Random Access Memory (ReRAM) \cite{park_neuron-inspired_2022} and Floating-Gate (FG) transistors. The FG device was fabricated using a 65nm CMOS process, and the ReRAM was fabricated by using  atomic layer deposition (ALD) and electron-beam (e-beam) evaporation. This study compares the performance of these two analog non-volatile devices on MNIST dataset \cite{noauthor_mnist_nodate} using a standard LeNet-5 architecture \cite{lecun_gradient-based_1998}, offering power, area, and accuracy estimates for each model and dataset. An overview of the proposed simulator is depicted in Figure \ref{fig:overview}.

\begin{figure*}[htb]
  \begin{center}
  \includegraphics[ width=0.8\textwidth]{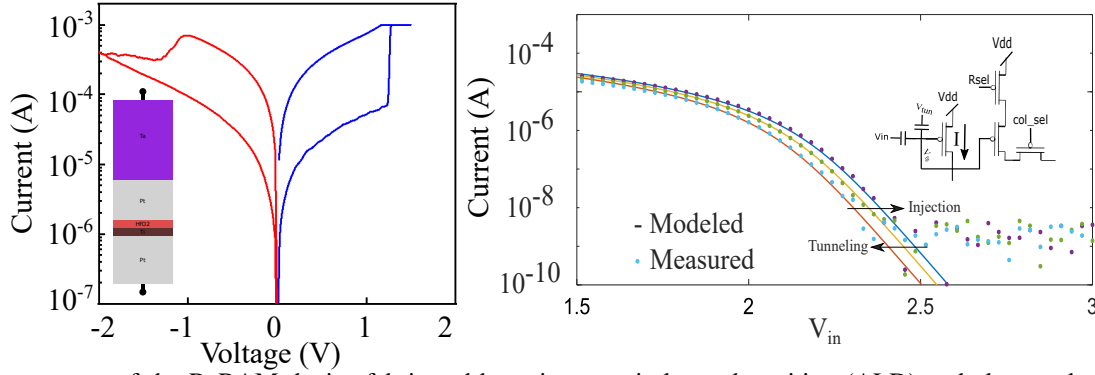}
  \vspace{-5mm}
  \end{center}
  \caption{Measurement of the ReRAM device fabricated by using  atomic layer deposition (ALD) and electron-beam (e-beam) evaporation on the left. Characterization of FG device fabricated in 65nm CMOS process on the right.}\label{fig:reram_and_fg}
  \vspace{-5mm}
\end{figure*}

\section{Non-volatile Devices} \label{sec:non_volatile}

\subsection{Resistive Random Access Memory }
Resistive Random Access Memory (ReRAM) is built using memristors or ReRAMs. ReRAMs are two terminal devices that can have different resistances programmed across the terminals. The devices can be constructed by sandwiching a metal oxide between two conducting electrodes, as shown in \ref{fig:reram_and_fg} \cite{park_neuron-inspired_2022}. The ReRAM used in this study is fabricated by using atomic layer deposition (ALD) and electron-beam (e-beam) evaporation. The ReRAM is constructed with a 50/25 nm Tantalum/Platinum top electrode, 5 nm of HfO2 switching oxide, and 5/30 nm Titanium/Platinum bottom electrode \cite{park_neuron-inspired_2022}.

Figure \ref{fig:reram_and_fg} shows the current vs. voltage characteristics of our ReRAM. Applying a positive voltage across the device (set) decreases its resistance. While applying a negative voltage across the device (reset) will increase its resistance\cite{park_neuron-inspired_2022}. These transitions are associated with the distribution of oxygen vacancies controllable through set (positive) and reset (negative) pulses. By applying voltages lower than set pulses, the resistance of the device can be measured without fluctuation. The length of the oxide minus the length of the filament is called the gap.

% \begin{figure}
%     \centering
%     \scalebox{0.32}{\includegraphics{Figures/reram_model.png}}
%     \caption{ (a) and (c) Schematic of the SET and RESET switching
%             processes and conductive filament configuration. (b) and (d) Simplified
%             model of the conductive filament at the SET and RESET states. LRS has
%             shorter (b) tunneling gap distance while HRS has (d) longer tunneling gap
%             distance\cite{7448912_compact_model}.}
%     \label{fig:rerammodel}
% \end{figure}

The current vs. voltage characteristics of the device is non-linear even when keeping the resistance state of the ReRAM device constant. In order to model this device during training and performance measurements we need to model the ReRAM's non-linear behavior. For this analysis, we used equation (\ref{equ:reramiv}) to model the devices I vs. V characteristics, where $V$ is the voltage across
\begin{equation}
    I(V) = I_0\exp{\bigg(-\frac{g}{g_0}\bigg)}\sinh{\bigg(\frac{V}{V_0}\bigg)}
    \label{equ:reramiv}
\end{equation}
the device, $I$ is the current through the device, $I_0$, $g_0$, and $V_0$ are calibration parameters obtained from measured data of a physical device and $g$ represents the gap between the end of the filament and the opposite electrode\cite{7448912_compact_model}.

\subsection{Floating-Gate Transistors}
Floating-gate transistors (FG) are a type of Field Effect Transistor (FET) with a floating gate that allows charge to be stored on it, resulting in a non-linear relationship between input voltage, stored charge, and output current. Figure \ref{fig:reram_and_fg}(b) shows a PMOS-FG transistor designed using the standard 65nm CMOS process, where the input is coupled through a capacitor. Charge can be modified on the floating gate by hot-electron injection or Fowler-Nordheim (FN) tunneling \cite{kim_floating-gate_2017,kim_integrated_2016}. Measured results for hot-electron injection and FN tunneling are presented in Figure \ref{fig:reram_and_fg}(b), showing how the drain current varies with these methods.

For hot-electron injection, a 4.5 V pulse is applied for 50 $\mu$s, resulting in an average threshold voltage change of 42.34mV, while a 5.8 V tunneling voltage is used to remove charges from the floating gate, leading to a threshold voltage change of 22.06mV. The floating node voltage depends on $V_{tun}$, $V_{DD}$, and $V_{out}$, and an EKV-derived transistor equation is used to model these relationships \cite{minch_translinear_1996,shah_temperature_2018}, as shown in equations \ref{Eq:Eq1} and \ref{Eq:Eq2}.
\vspace{-3mm}

\begin{equation}\label{Eq:Eq1}
\begin{aligned}
I_{synapse}=\\I_{th_{pmos}}(\ln^{2(1+{e^{{(\kappa(V_{DD}-V_{FG}-V_{TP})+\sigma(V_{DD}-V_d))}/2U_T}})} \\
    -\ln^{2(1+{e^{{(\kappa(V_{DD}-V_{FG}-V_{TP} )-(V_{DD}-V_d )}/2U_T}} )})    
\end{aligned}
\end{equation}
\vspace{-3mm}
\begin{equation}\label{Eq:Eq2}
\begin{aligned}
    V_{FG} \propto V_{FG_{0}} + C_{in} \times V_{in}/C_{T} + C_{gdo} \times V_d/ C_{T} \\
    + C_{gso}\times V_s /C_{T} + C_{tun}\times V_{tun}/C_{T}
\end{aligned}
\end{equation}

Equation \ref{Eq:Eq2} describes the total floating-gate voltage ($V_{FG}$), which is proportional to the floating gate charge ($Q_{FG}$), input ($V_{in}$), drain voltage ($V_{d}$), source voltage ($V_{s}$), and tunneling voltage ($V_{tun}$). In this equation, $C_{in}$ represents the input capacitance, while $C_{T}$ denotes the total capacitance on the floating node, including input capacitance, tunneling capacitance, oxide capacitance, and overlap capacitances of source and drain regions. The input and tunneling capacitance are implemented using Metal Oxide Semiconductor (MOS) caps with dimensions of $W\times L=3\mu m^2$ and $W\times L=0.16\mu m^2$, respectively. The study characterized FG transistors in a 65nm process and performed a fit using equation \ref{Eq:Eq1} to determine these parameters. The Python modeling employs the values derived from the fit, and Figure \ref{fig:reram_and_fg}(b) depicts the difference between the measured data and the EKV-derived model.

\section{Overall Architecture} \label{sec:overallarch}

\begin{figure}
    \centering
    \scalebox{0.2}{\includegraphics{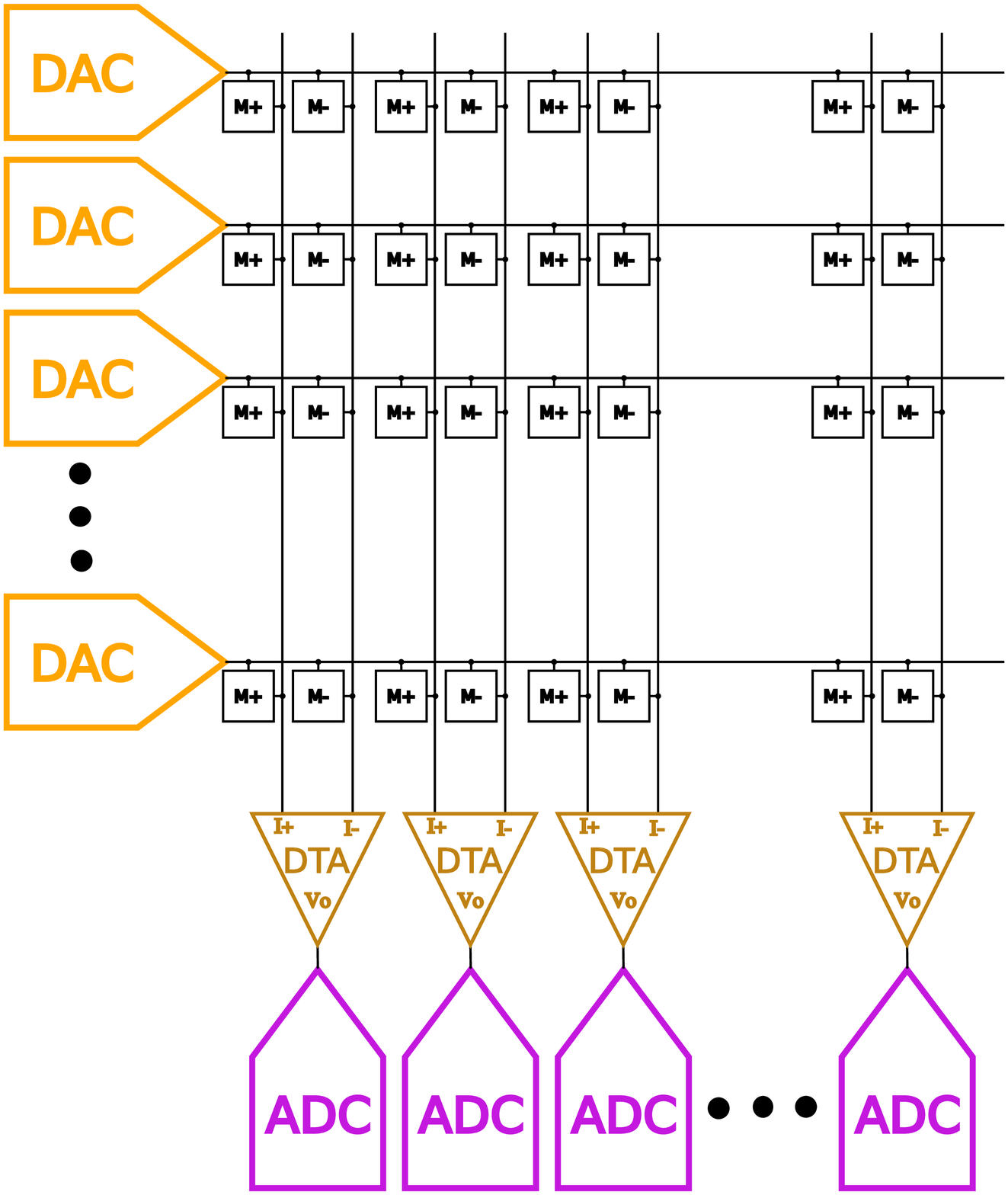}}
    \caption{ Overall Architecture for performing inference using the non-volatile memory. The DACs encode the inputs as an analog voltage, the memory elements output a current based on the DAC line voltage and stored value, the DTA outputs the difference in current as a voltage (with a fixed gain), and finally, the ADC converts the voltage back into a digital value as the output.}
    \label{fig:overallarch}
     \vspace{-5mm}
\end{figure}

Figure \ref{fig:overallarch} shows the non-volatile crossbar array for matrix-vector multiplication (MVM), consisting of four main elements: digital-to-analog converters (DACs), non-volatile memory devices (M+/-), differential transimpedance amplifiers (DTAs), and analog-to-digital converters. The matrix values are stored in the memory elements, while the input vector is encoded as voltages by the DACs. Each memory element outputs a current on its corresponding bit-line based on the voltage on its select-line (produced by the DAC) and the stored current value. KCL ensures that the currents flowing into the DTAs are the sum of the bit-line currents. To handle negative matrix values, each value is represented by a pair of memory elements, with M- producing a larger current than M+ for negative values. The DTAs subtract the currents (I+ - I-) and multiply by a fixed gain, resulting in the negative weights reducing the magnitude of the final voltage (Vo). Finally, the ADC converts the voltage produced by the DTA back into a digital value.

\vspace{-2mm}
\subsection{ReRAM Memory Element}\vspace{-2mm}
Based on equation (\ref{equ:reramiv}) we can see that the current through the ReRAM device depends on the state of the device $g$ and the voltage across it $V$. To utilize the ReRAM in MVM we encode the matrix values as a gap distance ($g$). The DACs encode the inputs as voltages on the bit-lines while the DTAs hold the voltage of the select-lines constant so the voltage across the ReRAM ($V$) is directly controlled by the bit-line voltage. Finally, the resulting current through the Re-RAMs is collected through the select-lines.
\vspace{-2mm}
\subsection{Floating Gate Memory Element}\vspace{-2mm}
Based on equation (\ref{Eq:Eq1}) and (\ref{Eq:Eq2}) we can see that the current through the FG depends on the floating node voltage $V_{FG_{0}}$ and the input $V_{in}$. To utilize the FG in MVM we encode the matrix values as floating node voltage ($V_{FG_{0}}$). The DACs encode the inputs as voltages on the bit-lines that represent $V_{in}$ in equation (\ref{Eq:Eq1}) and (\ref{Eq:Eq2}). The source of the FG is connected to $V_{DD}$ so $V_{S} = V_{DD}$ and the drain or the FG is connected to the select-line, which has its voltage held constant by the DTA. This allows ${V_{d}}$ and ${V_{s}}$ to remain constant and the select-line to collect the current of all the FGs connected to it.

\begin{table*}
\caption{LeNet-5 Training Results}
\label{tab:results}
\begin{tabularx}{\textwidth}{@{} l *{8}{c} c @{}}
%\begin{tabularx}{\textwidth}{c  c  c  c  c  c  c  c}
\toprule
\toprule
&Memory   
& IO & Weight & Peak Layer & Avg Layer & Peak Neuron 
& Avg Neuron & Overall \\ 
& Type & Quantization & Quantization & Power (W) & Power (W) & Current (mA) & Current (mA) & Accuracy (\%) \\

\midrule
\midrule
& ReRAM & $8$ bits & $36$ levels & $1.034$ & $0.24968$ & $0.1$ & $0.08$ & $97$\\
& & & & & $+ 0.454$ (DAC/ADC) & $+0.288$ (DAC/ADC) & & \\
\midrule
& FG & $8$ bits & $256$ levels & $0.0103$ & $0.0024$ & $0.03$ & $0.0015$ & $97$\\
& & & & & $+ 0.485$ (DAC/ADC) & $+0.288$ (DAC/ADC) & & \\
\bottomrule
\bottomrule
\end{tabularx}
\vspace{-5mm}
\end{table*}
\vspace{-2mm}

\subsection{DTA and ADCs}
ADCs, DACs and DTAs have a limited input and output dynamic range that limits the possible voltage and current values. There are two places where limits are imposed on the voltages and currents. One limit is imposed by the maximum amount of current that the DTA can sink from each of the select lines. The following equation $I_{DTA} = I_{max}*tanh(I_{sl}/I_{max})$ governs the current limit that is imposed during training and inference, where $I_{DTA}$ is the current that the DTA does sink on each bit-line, ${I_{max}}$ is the maximum current the DTA can sink (defined as a constant), and $I_{sl}$ is the current that would flow out of the bit-line if the DTA had no limit. The current limit was set to $0.1$ mA for ReRAMs and $1.0$ mA for FGs. These values were chosen based on the overall output current with respect to the input voltage of the devices.

The output voltage of the DTA is also limited by the power supply rails. The following equation $V_{ADC} = V_{max}*tanh(relu*(G*(I_{+} - I_{-})/V_{max}))$ is used to model the voltage limit during inference and training, where $V_{ADC}$ is the voltage supplied to the ADC for conversion, $V_{max}$ is the max output voltage of the DTA which was set to 0.5 V and 0.6 V for ReRAM and FGs, $I_+$ is the output from the equation that defines $I_{DTA}$ for the positive weights, $I_-$ is the output from the same equation for the negative weights, and $G$ is the fixed gain of the DTA. The limit imposed on the output voltage also serves as the activation function for the CNN. As this MVM architecture can only support positive input values the $relu$ function is used to model that.

\begin{figure}
    \centering
    \scalebox{0.4
    }{\includegraphics{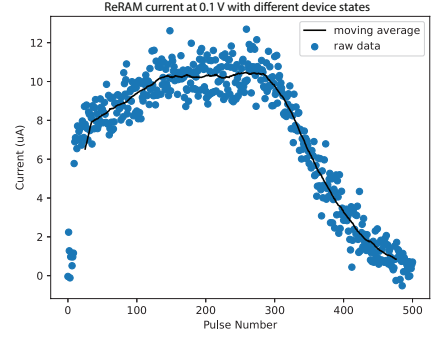}}
    \caption{Current measurements at 0.1 V with ReRAM device programmed at different states. \cite{park_neuron-inspired_2022}}
    \label{fig:reramsstates}
    \vspace{-5mm}
\end{figure}

\section{Learning Algorithms}
The non-linear behavior of the devices described in the previous section is embedded as part of the Python simulation infrastructure. Additionally, we use Stochastic Gradient Descent (SGD) to update the parameters of the convolution and fully connected layers model. These parameters correspond to a specific device property and the learning algorithm directly updates them. Additionally, we impose limits on the voltage and current to simulate saturation and finite dynamic range of the data converters. 

The current through the ReRAM device is modeled using equation \ref{equ:reramiv}, where the gap distance represents the weight stored in the ReRAM device. The weights are stored as a value from -1 to 1 but get inversely mapped to the minimum and maximum gap distance before being used in equation \ref{equ:reramiv}. The FG device used as a non-volatile analog device provides an output current that is non-linearly dependent on input voltage and stored charge as shown in equation \ref{Eq:Eq1}. The learning algorithm directly updates the $V_{FG0}$ shown in equation \ref{Eq:Eq2}. Additionally, we limit the input voltages between 0.2 to 0.6 volts during training and inference. 

%\subsection{Extra Hyperparamaters}
%Besides the standard hyperparameters that are used in CNNs, this architecture introduces $I_{max}$ from $I_{DTA}$ equation  and both $V_{max}$ and $G$ from equation (\ref{equ:vsat}). $G$ is adjusted to obtain reasonable training accuracy while also having the largest possible value. The larger this parameter the less current will need to flow to produce the same output and thus resulting in less total power consumption.

%$V_{max}$ controls what the input to a layer will be. Therefore it is chosen so that the voltage levels are large enough to avoid the need for high-precision DACs.

%Finally $I_{max}$ is chosen to reduce the overall power consumption and need for large transistors in the DACs and DTAs.

\section{Hardware Constraints and Simulator Results}
\vspace{-1mm}
Table \ref{tab:results} provides a summary of the results obtained for the two analog non-volatile devices. To accurately model the quantization observed in the measured devices, the study performed a quantization on the inputs, weights, and outputs of each layer, and computed the overall accuracy during testing. To simulate the dynamic range of digital-to-analog converters (DACs) and analog-to-digital converters (ADCs), linear quantization was applied to the layer inputs and outputs. However, the weights programmed into the floating gates (FGs) were quantized linearly, since they are represented as floating node voltages. For Resistive Random Access Memories (ReRAMs), a moving average over a set of resistance states was used to quantize the trained weights from a measured device, as the distribution of resistance states is not linear. 

%Secondly, hyperparameter selection was carried out, where $V_{max}$ was set to $500$ mV and $600$ mV for ReRAMs and FGs, respectively. The DTA gain was set to $2.903 \times 10^5 \Omega$ and $6 \times 10^6 \Omega$ for ReRAMs and FGs, respectively. The current limit was set to $0.1$ mA for ReRAMs and $1.0$ mA for FGs. These values were chosen based on the overall output current with respect to the input voltage of the devices.

The power consumption shown in Table \ref{tab:results} also includes power from an 8-bit ADC and 8-bit DAC with a dynamic range of 0.2 V to 0.8 V. For this initial study we simulated the ADC power draw in 130nm (where ReRAM devices are available) and 65nm CMOS process (where our current FG devices are fabricated). The average power draw of a single ADC was $9.38\mu W$ in both processes. A current steering DAC was also simulated using the same technology nodes which resulted in a power consumption of $400\mu W$. For this analysis, the ADC power consumption of a layer was calculated to be the ADC power multiplied by the number of outputs for that layer. For the DAC power consumption of a layer, the power consumption of a single DAC is multiplied by the number of inputs for that layer.

The area for a pair of ReRAM devices (with an access transistor) is $8.64\mu m^2$ and the area for a FG device pair is $78.72 \mu m^2$. For this study each trainable weight was stored in a single device pair. Each DAC has an area of $25600 \mu m^2$ and each ADC has an area of $6681.1 \mu m^2$.

% \begin{center}
% \begin{tabular}{|c | c |}
% \hline
% Layer & ADC + DAC Average Power (W) \\
% \hline
% Conv2d & 0.454 \\
% Conv2d & 0.485 \\
% Conv2d & 0.161 \\
% Dense  & 0.049 \\
% Dense  & 0.034 \\
% \hline
% \end{tabular} \
% \end{center}
%For ADC: 9.38uA what is VDD (using 1V for now)?
%For DAC: 400uW.

Further, to obtain the overall power and current dissipation the simulator enables calculating the average and peak power and current for all neurons in each layer. The layer power was calculated by summing the currents into each DTA and multiplying by $V_{DD}$. The inference simulation performed in this analysis did not account for the rise and fall time for the DACs or acquisition time for the ADCs. Therefore, the peak power reported is the maximum power that the layer would dissipate given a set of inputs and weights. The final power average was calculated by taking the average over all the averages calculated for each layer, and the final peak power was calculated by taking the peak power out of each layer's peak power.

% input: torch.Size([32, 1, 32, 32])        conv1: 32 * 32  -> 6 * 28 * 28    1024 -> 4704  4816896
% output: torch.Size([32, 6, 28, 28])      

% intput: torch.Size([32, 6, 14, 14])       conv2: 6 * 14 * 14 -> 16 * 10 * 10  1176 -> 1600  1881600
% output: torch.Size([32, 16, 10, 10])

% input: torch.Size([32, 16, 5, 5])         conv3: 16 * 5 * 5 -> 120           400 -> 120 48000
% output: torch.Size([32, 120, 1, 1])

% input: torch.Size([32, 120])              fc1: 120 -> 84                     120 -> 84  10080
% output: torch.Size([32, 84])              

% intput: torch.Size([32, 84])              fc2: 84 -> 10                     84 -> 10 840
%output: torch.Size([32, 10])

%FG area:
% (4816896 + 1881600 + 48000 + 10080 + 840) * 15.87 * 2.48 um^2

\section{Discussion}

Overall both the ReRAM and FG model was able to be trained to perform inferences on the MNIST dataset with models that closely match the device characteristics \cite{noauthor_mnist_nodate}. Table \ref{tab:results} shows the trade-offs between the ReRAM and FG memories. The overall area for the ReRAMs was significantly lower but the overall power was significantly higher. This is because FGs produce a lower current given the same input voltage as compared to the ReRAM devices. While the ReRAM devices are fabricated in the metal layers of the process and only require one transistor.

With the simulated neuron currents, we can make better-informed decisions on transistor sizing for the DACs and DTAs. Also by comparing the power from the tile current with the power from the DACs + ADCs we can make informed decision on the sizing of tiles \cite{tiled_reram}. The more inputs a neuron has the more current will flow through the neuron. The ReRAM neuron currents were higher meaning that smaller tiles would need to be used to maintain accuracy and reasonable transistor sizing on models with a larger number of  parameters (weights and biases).

Another factor that can go into tile sizing is the ratio between the power due to the tile current and the DAC + ADC power. As the tile grows larger the DAC + ADC power grows linearly, while the power from tile current grows quadratically. It is important that the tile size is large enough that the power overhead from the DACs and ADCs is small compared to the power from the neuron currents. When optimizing for area a similar relationship applies to the area of the DACs and ADCs v.s. the area of the memory devices.

In this analysis we sized the tiles such that they match the layers in the LeNet-5 architecture. Based on the various tile sizes we can see that peak power layer for the ReRAMs had an DAC + ADC overhead of $0.454 W$ which is less than the average power from the ReRAM's tile current ($0.25 W$). However the peak power layer overhead for the FGs ($0.485W$) was much larger then the average power from the FG's tile current ($0.0024 W$). This means that the sizing for the ReRAM tile is reasonable (for the peak power layer) but could be made larger for the FGs. The smallest layer in the architecture had a DAC + ADC overhead of $0.034 W$ which was much higher than the tile current power for that specific layer ($0.003W$ for ReRAMs, and $1.4\mu W$ for the FGs).

%\newpage
% \begin{figure*}[htb]
%  \begin{center}    
%  \epsfig{file=Figures/Figure2_FG.eps, width=\textwidth}
%  \end{center}
%  \vspace{-5mm}
%  \caption{ }\label{fig:results}
%  \vspace{-5mm}
%\end{figure*}

%\section{Acknowledgements}
%This study was partly funded by a grant agency (xyyx) and partly through a seed grant (zzss). 

%\clearpage
\bibliographystyle{IEEEtran}
\bibliography{./references1.bib, references2.bib}

\begin{thebibliography}{10}
\providecommand{\url}[1]{#1}
\csname url@samestyle\endcsname
\providecommand{\newblock}{\relax}
\providecommand{\bibinfo}[2]{#2}
\providecommand{\BIBentrySTDinterwordspacing}{\spaceskip=0pt\relax}
\providecommand{\BIBentryALTinterwordstretchfactor}{4}
\providecommand{\BIBentryALTinterwordspacing}{\spaceskip=\fontdimen2\font plus
\BIBentryALTinterwordstretchfactor\fontdimen3\font minus
  \fontdimen4\font\relax}
\providecommand{\BIBforeignlanguage}[2]{{%
\expandafter\ifx\csname l@#1\endcsname\relax
\typeout{** WARNING: IEEEtran.bst: No hyphenation pattern has been}%
\typeout{** loaded for the language `#1'. Using the pattern for}%
\typeout{** the default language instead.}%
\else
\language=\csname l@#1\endcsname
\fi
#2}}
\providecommand{\BIBdecl}{\relax}
\BIBdecl

\bibitem{mcguire_minimal_2019}
\BIBentryALTinterwordspacing
K.~N. McGuire, C.~De~Wagter, K.~Tuyls, H.~J. Kappen, and G.~C. H.~E. de~Croon,
  ``\BIBforeignlanguage{en}{Minimal navigation solution for a swarm of tiny
  flying robots to explore an unknown environment},''
  \emph{\BIBforeignlanguage{en}{Science Robotics}}, vol.~4, no.~35, p.
  eaaw9710, Oct. 2019. [Online]. Available:
  \url{https://robotics.sciencemag.org/lookup/doi/10.1126/scirobotics.aaw9710}
\BIBentrySTDinterwordspacing

\bibitem{olsson_event_2016}
R.~H. Olsson, R.~B. Bogoslovov, and C.~Gordon, ``Event driven persistent
  sensing: {Overcoming} the energy and lifetime limitations in unattended
  wireless sensors,'' in \emph{2016 {IEEE} {SENSORS}}, Oct. 2016, pp. 1--3.

\bibitem{han_learning_nodate}
S.~Han, J.~Pool, J.~Tran, and W.~Dally, ``\BIBforeignlanguage{en}{Learning both
  {Weights} and {Connections} for {Efficient} {Neural} {Network}},'' p.~9.

\bibitem{sze_designing_2017}
V.~Sze, ``Designing {Hardware} for {Machine} {Learning}: {The} {Important}
  {Role} {Played} by {Circuit} {Designers},'' \emph{IEEE Solid-State Circuits
  Magazine}, vol.~9, no.~4, pp. 46--54, 2017, number: 4 Conference Name: IEEE
  Solid-State Circuits Magazine.

\bibitem{chowdhury_hardware_2022}
S.~Chowdhury and S.~Shah, ``Hardware aware modeling of mixed-signal spiking
  neural network,'' \emph{IEEE NEWCAS}, 2022.

\bibitem{ankit_puma_2019}
\BIBentryALTinterwordspacing
A.~Ankit, I.~E. Hajj, S.~R. Chalamalasetti, G.~Ndu, M.~Foltin, R.~S. Williams,
  P.~Faraboschi, W.-m. Hwu, J.~P. Strachan, K.~Roy, and D.~S. Milojicic,
  ``{PUMA}: {A} {Programmable} {Ultra}-efficient {Memristor}-based
  {Accelerator} for {Machine} {Learning} {Inference},'' \emph{arXiv:1901.10351
  [cs]}, Jan. 2019, arXiv: 1901.10351. [Online]. Available:
  \url{http://arxiv.org/abs/1901.10351}
\BIBentrySTDinterwordspacing

\bibitem{ali_imac_2020}
M.~Ali, A.~Jaiswal, S.~Kodge, A.~Agrawal, I.~Chakraborty, and K.~Roy, ``{IMAC}:
  {In}-{Memory} {Multi}-{Bit} {Multiplication} and {ACcumulation} in {6T}
  {SRAM} {Array},'' \emph{IEEE Transactions on Circuits and Systems I: Regular
  Papers}, vol.~67, no.~8, pp. 2521--2531, Aug. 2020, conference Name: IEEE
  Transactions on Circuits and Systems I: Regular Papers.

\bibitem{mutlu_processing_2019}
\BIBentryALTinterwordspacing
O.~Mutlu, S.~Ghose, J.~Gómez-Luna, and R.~Ausavarungnirun,
  ``\BIBforeignlanguage{en}{Processing data where it makes sense: {Enabling}
  in-memory computation},'' \emph{\BIBforeignlanguage{en}{Microprocessors and
  Microsystems}}, vol.~67, pp. 28--41, Jun. 2019. [Online]. Available:
  \url{https://www.sciencedirect.com/science/article/pii/S0141933118302291}
\BIBentrySTDinterwordspacing

\bibitem{fan_leveraging_2017}
D.~Fan, Z.~He, and S.~Angizi, ``Leveraging spintronic devices for ultra-low
  power in-memory computing: {Logic} and neural network,'' in \emph{2017 {IEEE}
  60th {International} {Midwest} {Symposium} on {Circuits} and {Systems}
  ({MWSCAS})}, Aug. 2017, pp. 1109--1112, iSSN: 1558-3899.

\bibitem{sengupta_hybrid_2016}
\BIBentryALTinterwordspacing
A.~Sengupta, A.~Banerjee, and K.~Roy, ``\BIBforeignlanguage{en}{Hybrid
  {Spintronic}-{CMOS} {Spiking} {Neural} {Network} with {On}-{Chip} {Learning}:
  {Devices}, {Circuits}, and {Systems}},''
  \emph{\BIBforeignlanguage{en}{Physical Review Applied}}, vol.~6, no.~6, p.
  064003, Dec. 2016. [Online]. Available:
  \url{https://link.aps.org/doi/10.1103/PhysRevApplied.6.064003}
\BIBentrySTDinterwordspacing

\bibitem{khaddam-aljameh_hermes_2021}
R.~Khaddam-Aljameh, M.~Stanisavljevic, J.~Fornt~Mas, G.~Karunaratne,
  M.~Braendli, F.~Liu, A.~Singh, S.~M. Müller, U.~Egger, A.~Petropoulos,
  T.~Antonakopoulos, K.~Brew, S.~Choi, I.~Ok, F.~L. Lie, N.~Saulnier, V.~Chan,
  I.~Ahsan, V.~Narayanan, S.~R. Nandakumar, M.~Le~Gallo, P.~A. Francese,
  A.~Sebastian, and E.~Eleftheriou, ``{HERMES} {Core} – {A} 14nm {CMOS} and
  {PCM}-based {In}-{Memory} {Compute} {Core} using an array of 300ps/{LSB}
  {Linearized} {CCO}-based {ADCs} and local digital processing,'' in \emph{2021
  {Symposium} on {VLSI} {Circuits}}, Jun. 2021, pp. 1--2, iSSN: 2158-5636.

\bibitem{tiled_reram}
Q.~Wang, X.~Wang, S.~H. Lee, F.-H. Meng, and W.~D. Lu, ``A deep neural network
  accelerator based on tiled rram architecture,'' in \emph{2019 IEEE
  International Electron Devices Meeting (IEDM)}, 2019, pp. 14.4.1--14.4.4.

\bibitem{park_neuron-inspired_2022}
\BIBentryALTinterwordspacing
M.~Park, Y.~Yuan, Y.~Baek, A.~H. Jones, N.~Lin, D.~Lee, H.~S. Lee, S.~Kim,
  J.~C. Campbell, and K.~Lee, ``\BIBforeignlanguage{en}{Neuron-{Inspired}
  {Time}-of-{Flight} {Sensing} via {Spike}-{Timing}-{Dependent} {Plasticity} of
  {Artificial} {Synapses}},'' \emph{\BIBforeignlanguage{en}{Advanced
  Intelligent Systems}}, vol.~4, no.~3, p. 2100159, 2022, \_eprint:
  https://onlinelibrary.wiley.com/doi/pdf/10.1002/aisy.202100159. [Online].
  Available:
  \url{https://onlinelibrary.wiley.com/doi/abs/10.1002/aisy.202100159}
\BIBentrySTDinterwordspacing

\bibitem{noauthor_mnist_nodate}
\BIBentryALTinterwordspacing
``{MNIST} handwritten digit database, {Yann} {LeCun}, {Corinna} {Cortes} and
  {Chris} {Burges}.'' [Online]. Available:
  \url{http://yann.lecun.com/exdb/mnist/}
\BIBentrySTDinterwordspacing

\bibitem{lecun_gradient-based_1998}
Y.~Lecun, ``\BIBforeignlanguage{en}{Gradient-{Based} {Learning} {Applied} to
  {Document} {Recognition}},'' \emph{\BIBforeignlanguage{en}{PROCEEDINGS OF THE
  IEEE}}, vol.~86, no.~11, 1998.

\bibitem{7448912_compact_model}
Z.~Jiang, Y.~Wu, S.~Yu, L.~Yang, K.~Song, Z.~Karim, and H.-S.~P. Wong, ``A
  compact model for metal–oxide resistive random access memory with
  experiment verification,'' \emph{IEEE Transactions on Electron Devices},
  vol.~63, no.~5, pp. 1884--1892, 2016.

\bibitem{kim_floating-gate_2017}
S.~Kim, S.~Shah, and J.~Hasler, ``Floating-gate {FPAA} calibration for analog
  system design and built-in self test.''\hskip 1em plus 0.5em minus
  0.4em\relax IEEE, 2017, pp. 1--4.

\bibitem{kim_integrated_2016}
S.~Kim, J.~Hasler, and S.~George, ``Integrated {Floating}-{Gate} {Programming}
  {Environment} for {System}-{Level} {ICs},'' \emph{IEEE Transactions on Very
  Large Scale Integration (VLSI) Systems}, vol.~24, no.~6, pp. 2244--2252, Jun.
  2016, conference Name: IEEE Transactions on Very Large Scale Integration
  (VLSI) Systems.

\bibitem{minch_translinear_1996}
\BIBentryALTinterwordspacing
B.~A. Minch, C.~Diorio, P.~Hasler, and C.~A. Mead,
  ``\BIBforeignlanguage{en}{Translinear circuits using subthreshold
  floating-gate {MOS} transistors},'' \emph{\BIBforeignlanguage{en}{Analog
  Integrated Circuits and Signal Processing}}, vol.~9, no.~2, pp. 167--179,
  Mar. 1996. [Online]. Available:
  \url{http://link.springer.com/10.1007/BF00166412}
\BIBentrySTDinterwordspacing

\bibitem{shah_temperature_2018}
S.~Shah, H.~Toreyin, J.~Hasler, and A.~Natarajan, ``Temperature {Sensitivity}
  and {Compensation} on a {Reconfigurable} {Platform},'' \emph{IEEE
  Transactions on Very Large Scale Integration (VLSI) Systems}, vol.~26, no.~3,
  pp. 604--607, 2018, publisher: IEEE.

\end{thebibliography}

\ifCLASSOPTIONcaptionsoff
  \newpage
\fi

\end{document}